# High-temperature Superconducting Oxide without Copper at Ambient Pressure

S. Lin Er Chow,[†] Zhaoyang Luo, A. Ariando[†]

*Department of Physics, Faculty of Science, National University of Singapore, Singapore 117551, Singapore*

[†]Corresponding author: le.chow@u.nus.edu, ariando@nus.edu.sg

**The discovery of superconductivity in the Ba-La-Cu-O system (the cuprate) at the 30 K range in 1986 marked a significant breakthrough, as it far exceeded the highest known critical temperature ($T_c$) at the time and surpassed the predicted 30 K limit, which was thought to be the maximum before phonon-mediated electron pairing would break down due to thermal excitation. Despite recent successful observations of superconductivity in nickel-oxide-based compounds (the nickelate), superconductivity above 30 K at ambient pressure in a system that is isostructural and isoelectronic to the cuprate but without copper has remained elusive. Here, we report a superconducting $T_c$ above 35 K under ambient pressure in hole-doped, late rare-earth infinite-layer nickel oxide (Sm-Eu-Ca-Sr)NiO$_2$ thin films. Electron microscopy reveals a small thickness of ~ 2 nm of infinite-layer phase stabilised at present, which indicates a higher temperature superconductivity should be observable in clean bulk crystals.**

A longstanding question in condensed matter is whether high-temperature (high-$T_c$) superconductivity above 30 K at ambient pressure is unique to quasi-two-dimensional copper oxide (cuprate) or ubiquitous among transition metal oxides in the periodic table of elements [1–4]. As one of the candidates that Bednorz and Müller attempted to synthesise for the realisation of high-$T_c$ superconducting oxides [5–7], intense efforts into the synthesis of nickel oxide (nickelate) have been undertaken for more than three decades [8–11]. However, only a few members of nickelate superconductors have been successfully synthesised to date, where superconductivity at ambient pressure is only observed in the $d^{9-x}$ doped infinite-layer (La/Pr/Nd)NiO$_2$ [12–16] and quintuple layer Nd$_6$Ni$_5$O$_{12}$ [17], while a high-pressure induced superconducting phase is recently observed in the $d^{7+x}$ nickelates [18]. The challenging material synthesis and the limited number of superconducting members in the nickelates family have been bottlenecking fundamental understanding of the correlated system and its superconducting pairing mechanism [19–23]. For example, high-field transport studies on the hole-doped (La/Pr/Nd)NiO$_2$ show a drastic difference in the scale of upper critical fields between various rare-earth nickelates, where a large Pauli-limit violation in all directions is observed in the Ca/Sr-doped LaNiO$_2$ [24,25], but absent in the Nd-nickelates counterpart [26]. Meanwhile, the low-temperature penetration depth and superfluid density studies suggest a nodal gap in the La/Pr-nickelates, but a nodeless gap in the Nd-nickelates [27,28]. Such an ambiguous distinction between different rare-earth nickelates begs the importance of discovering a new nickelate superconductor beyond the La/Pr/Nd rare earth elements.

In the high-$T_c$ cuprates, the largest rare-earth ion La$_{2-x}$Ce$_x$CuO$_{4+\delta}$ has $T_c \sim 30$ K but Eu$_{2-x}$Ce$_x$CuO$_{4+\delta}$ has a lower $T_c \sim 10$ K [29,30]. For the infinite-layer nickelates, their three-dimensional electronic structures mark the first deviation from cuprates' quasi-two-dimensional single-band picture [7,31–35]. It is, therefore, critical to understand the rare-earth dependence superconducting $T_c$, phase diagram, dimensionality, order parameter, Fermi surfaces, symmetries, and competing orderings of the strongly-correlated nickelates [21,36–39].

Samarium-nickelate is the next candidate to expand the rare-earth infinite-layer nickelates families. Considering Sr-doped (La/Pr/Nd)NiO$_2$ are all superconducting at 20% Sr doping, 0.2 Sr-doped SmNiO$_2$ is expectedly superconducting as well. We observed a metallic behaviour (with small resistivity upturn below 50 K) and low resistivity (~1 m$\Omega\cdot$cm ) comparable to the superconducting hole-doped (Nd/Pr/La)NiO$_2$ [12,13,40], but no sign of superconducting transition down to 2 K [41]. This observation should not be considered conclusive. We then attempted to hole dope with a smaller cation, such as Ca$^{2+}$ or Eu$^{2+/3+}$, following our previous work on the dimensionality manipulation of superconductivity in the infinite-layer nickelates [42]. In addition, we are encouraged by the possibility of better stabilising the hole-doped perovskite SmNiO$_3$ through mix-valent Eu$^{2+/3+}$ [16]. With these considerations, we focus on synthesising doped Sm$_{1-x-y-z}$Eu$_x$Ca$_y$Sr$_z$NiO$_2$ (SECS) infinite-layer nickelates. The crystal structure of this superconducting candidate is schematically illustrated in **Fig. 1a**.

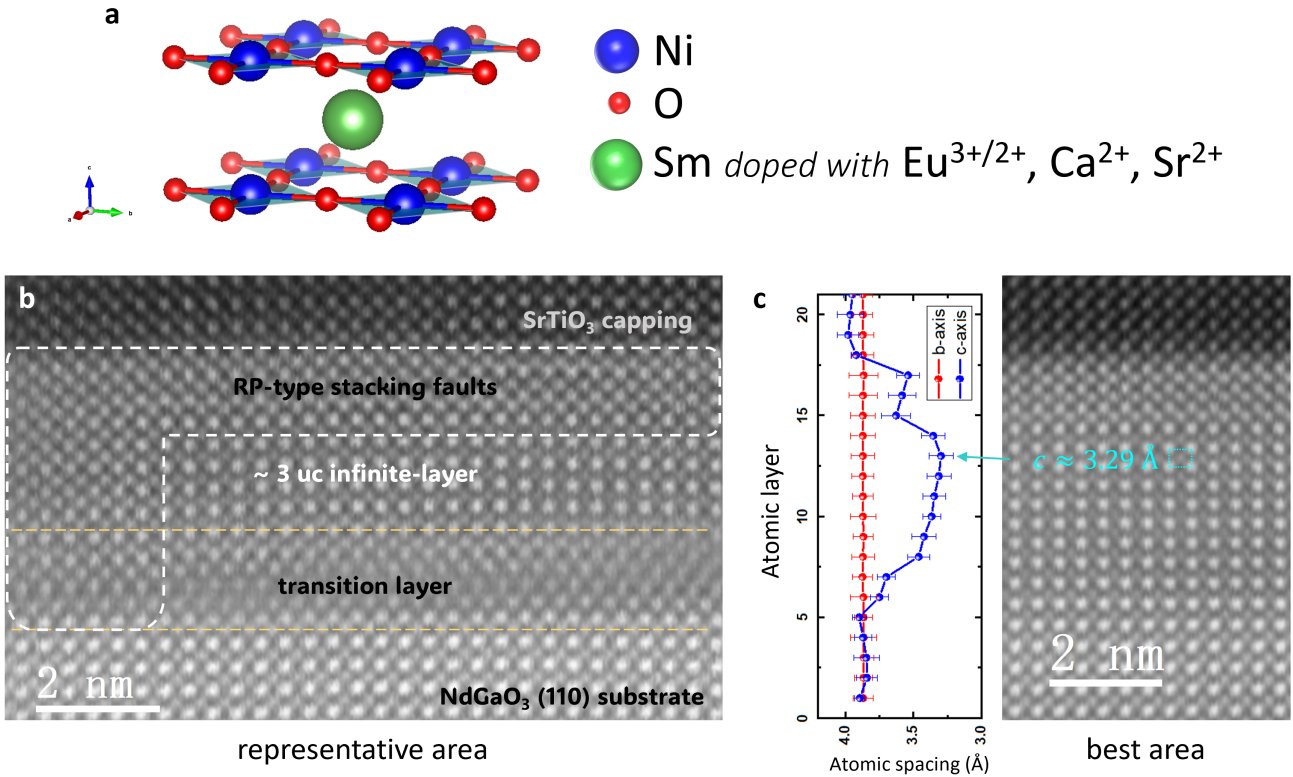

FIG. 1. Structural information on the superconducting Sm$_{1-x-y-z}$Eu$_x$Ca$_y$Sr$_z$NiO$_2$ (SECS) thin film. (a) The schematic diagram for the crystal structure of infinite-layer nickelates SmNiO$_2$ doped with Eu$^{3+/2+}$, Ca$^{2+}$, Sr$^{2+}$. (b) Cross-

sectional scanning transmission electron microscopy (STEM) high-angle annular dark field (HAADF) image of a representative superconducting $Sm_{0.74}Eu_{0.06}Ca_{0.01}Sr_{0.19}NiO_2$ thin film. (c) Calculation of the atomic spacing in the best crystalline area.

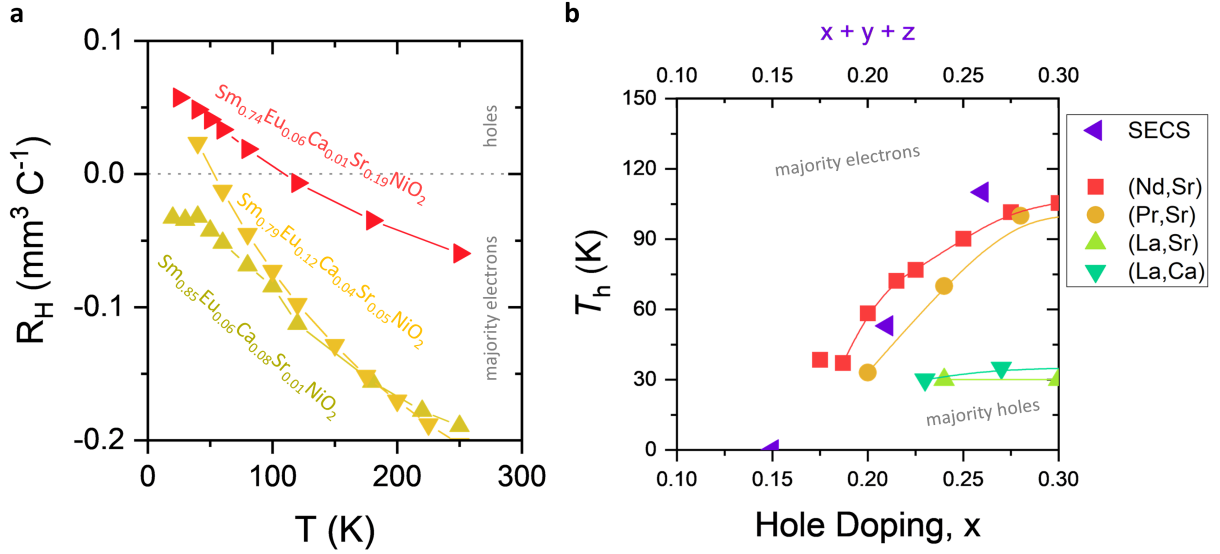

FIG. 2. Normal state transport characteristics of the $Sm_{1-x-y-z}Eu_xCa_ySr_zNiO_2$ (SECS) thin films. (a) Hall coefficients vs temperature $R_H(T)$ plot. (b) Phase diagram of various infinite-layer nickelates $(R, A)NiO_2$ showing $R_H$ sign crossover temperature $T_h$ as a function of hole doping, adapted from ref. [13,15,43,44].

**Fig. 1** shows the cross-sectional scanning transmission electron microscopy of a representative superconducting infinite-layer nickelates thin film $Sm_{0.74}Eu_{0.06}Ca_{0.01}Sr_{0.19}NiO_2$ grown on $NdGaO_3(110)$. The $c$-axis lattice constant of the infinite-layer phase is calculated to be ~ 3.29 Å, which agrees well with the theoretical values for $P4/mmm$ space group [45]. A transition layer up to ~ 4 uc can be observed at the film-substrate interface [46] in most parts of the film (**Fig. 1b**) [41]. In addition, large areas of Ruddlesden–Popper (RP) type stacking faults are present throughout the film, suppressing the maximum achievable infinite-layer phase to ~ 3 uc – 5 uc (**Fig. 1b-c**). We note that this thickness is much smaller than those for hole-doped $(Nd/Pr/La)NiO_2$, where the infinite-layer phase can be stabilised up to ~ 15 uc – 30 uc [16,22,27,43,47]. The high defect density is reflected in the magnetoresistance of the infinite-layer nickelates [41], where a small negative magnetoresistance is observed. The perovskite phase is mostly insulating below room temperature [41], similar to the previous reports on undoped perovskite $SmNiO_3$ [46]. **Fig. 2** shows the Hall transport characteristics of the SECS infinite-layer thin films. The overdoped $x + y + z = 0.26$ SECS shows a sign change in Hall coefficients $R_H$ from negative at high temperatures to positive below $T_h = 110$ K (**Fig. 2a**). This sign-change temperature $T_h$ is slightly higher than those of overdoped (Nd/Pr/La)$NiO_2$ [13,15,44,48] (**Fig. 2b**), suggesting smaller rare-earth electron pockets in the $SmNiO_2$. On the other hand, underdoped $Sm_{0.85}Eu_{0.06}Ca_{0.08}Sr_{0.01}NiO_2$ has negative $R_H$ and majority electrons charge carriers at all temperatures measured, which is consistent with the underdoped infinite-layer nickelate of earlier rare-earth elements. The underdoped $Sm_{0.85}Eu_{0.06}Ca_{0.08}Sr_{0.01}NiO_2$ has a lower $T_c \approx 7$ K [41]. We note that (for convenience, $T_c$ refers to $T_{c,onset}$ throughout this manuscript) $T_c$ value in this report should not be considered definitive, as a significantly higher (increase up to $\times 2$) $T_c$ has shown to be achievable through synthesis advances in the hole-doped (Nd/Pr/La)$NiO_2$ [43,47].

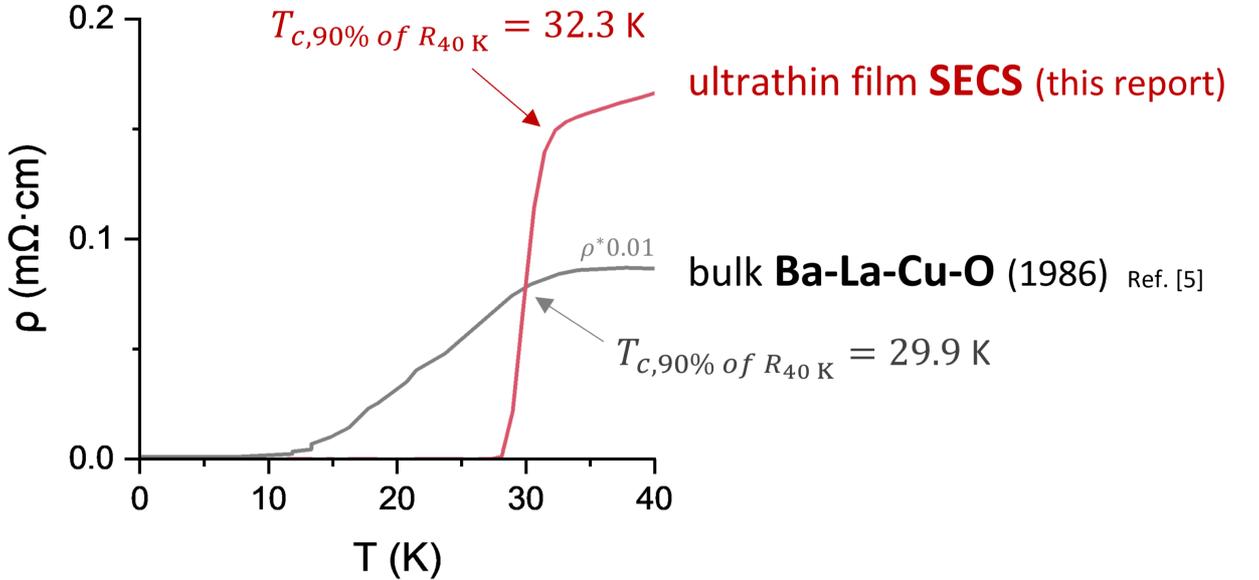

FIG. 3. Superconductivity in the $Sm_{0.79}Eu_{0.12}Ca_{0.04}Sr_{0.05}NiO_2$ (SECS) thin film, in comparison with the first cuprate Ba-La-Cu-O (1986). Resistivity *vs* temperature $\rho(T)$ curve of Ba-La-Cu-O is extracted from Ref. [5].

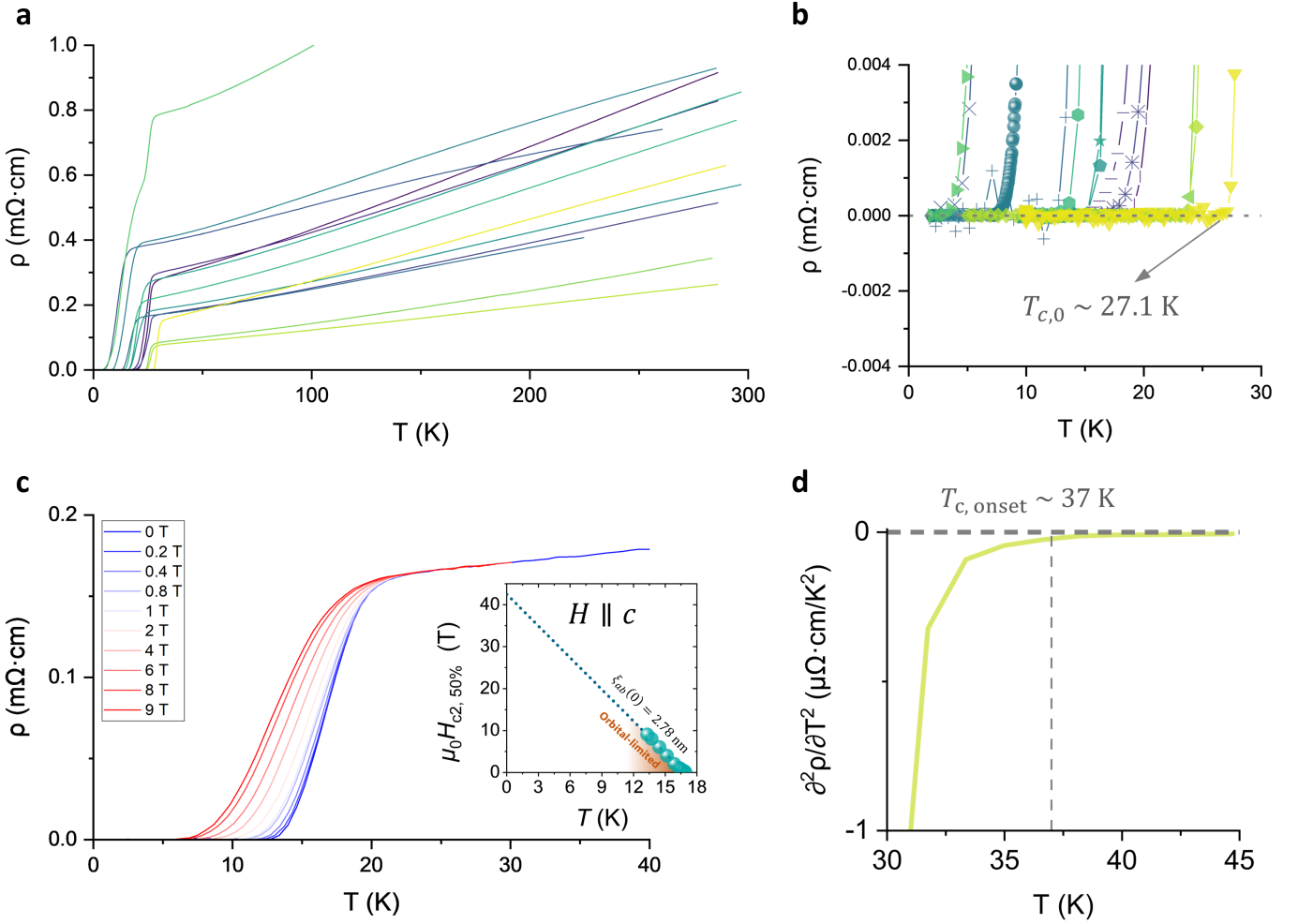

FIG. 4. Superconductivity in the $Sm_{1-x-y-z}Eu_xCa_ySr_zNiO_2$ (SECS) thin films. (a) Resistivity vs temperature $\rho(T)$ curves of several superconducting SECS samples. (b) Magnification of the $\rho(T)$ curves at zero resistivity. (c) Magnetic field dependence $\rho(T)$ for a representative SECS sample. The field was applied along the perpendicular direction ($H \parallel c$). Inset shows the upper critical field $H_{c2} - T$ taken at 50% transition criteria, with a linear fit. (d) 2$^{nd}$ order derivative $\frac{\partial^2 \rho}{\partial T^2}(T)$ of the $\rho(T)$ curve. The onset of superconductivity is defined as the temperature where $\frac{\partial^2 \rho}{\partial T^2}$ drops below zero ($\frac{\partial^2 \rho}{\partial T^2} < 0$) [43].

**Figs. 3-4** show the resistivities versus temperature plots of the superconducting infinite-layer $Sm_{1-x-y-z}Eu_xCa_ySr_zNiO_2$ (SECS). We can reproduce superconductivity in more than ten samples, where a zero-resistance state is evidently present in **Fig. 4b**. The highest $T_{c,0} \approx 27.1$ K is the first superconducting nickelate which zero resistance state temperature surpasses the liquid hydrogen boiling point, while a higher $T_c$ remains promising for SECS, given much room for sample quality improvement. **Fig. 4d** shows

the highest onset of superconductivity observed at $T_{c,onset} \sim 37$ K, where the onset of superconducting transition is taken as the temperature where 2nd order derivative $\frac{\partial^2 \rho}{\partial T^2}$ drops below zero $\left(\frac{\partial^2 \rho}{\partial T^2} < 0\right)$, following the convention used in nickelates [43]. Using other definitions of the onset of superconductivity would yield a $T_c$ value above 30 K as well, comparable to the first cuprate Ba-La-Cu-O (**Fig. 3**) [5]. The suppression of superconductivity by applied magnetic fields is representatively shown in **Fig. 4c** and Ref. [41], supporting the observation of superconductivity in the hole-doped SmNiO$_2$. We extract the upper critical fields along the perpendicular to $ab$ plane ($c$-axis) direction using 50% transition criteria, $\mu_0 H_{c2,50\%}$ (**Fig. 4c** inset). We fit the $\mu_0 H_{c2,50\%}(T)$ to the linearised Ginzburg-Landau equation: $\mu_0 H_{c2,50\%}(T) = \frac{\Phi_0}{2\pi \xi_{ab}^2(0)} \left(1 - \frac{T}{T_c}\right)$ where $\Phi_0$ is the flux quantum and $\xi_{ab}(0)$ is the zero-temperature Ginzburg-Landau coherence length, $\xi_{ab}(0) = 2.78$ nm for this sample. A shorter coherence length is expected for the higher $T_c$ sample [41]. The linear dependence $\mu_0 H_{c2,50\%}(T)$ suggests a dominant orbital pair-breaking mechanism within this temperature range.

**Fig. 5** summarizes all the discovered superconducting nickelates at ambient pressure, where the maximum onset $T_c$ is observed at $< 20$ K [12–15], and up to $\sim 25$ K for optimally grown strained nickelates (hole-doped NdNiO$_2$ on NdGaO$_3$ [49] and LSAT [16,43] substrates). The newly discovered SECS grown on NdGaO$_3$ (NGO) would have a small strain of $\sim 0.26\%$ if epitaxially strained, which should be insignificant for modulating $T_c$, as shown in the recent freestanding works [49,50]. The realisation of superconductivity for infinite-layer nickelates grown directly on NGO substrate here also excludes the role of TiO$_2$ interface [12] in achieving superconductivity in doped RNiO$_2$ grown on SrTiO$_3$ substrate or buffer layer [49,50], and LSAT substrate [16,43]. On the other hand, large regions of transition layer at the SECS/NGO interface and Ruddlesden-Popper type stacking faults bottleneck the thickness of the infinite-layer phase and maximum $T_c$ of SECS. At present material quality, only $\sim 3 - 5$ unit cells ($< 2$ nm) thickness of superconducting infinite-layer phase achieved, which should severely limit the superconducting $T_c$ due to confinement effect, similar to those of few nm YBCO or Nb thin films [51]. For the quasi-2D cuprate YBCO, a ~ 2.3 nm thick film has only ~ 56% of bulk $T_c$ [52]. Considering the more 3D electronic structure of infinite-layer nickelates as compared to the cuprates [32,53–55], it can be extrapolated that their intrinsic bulk $T_c \geq$ 37 K/0.56 ≈ 66 K. In addition, electron microscopy reveals that most parts of the lattice are oxygen off stoichiometric $ANiO_{2+\delta}$ where apical oxygen is not completely removed [41], which drastically lowers the superconducting $T_c$ of SECS to the present value. It was demonstrated in the Sr-doped $LaNiO_2$ where $T_c$ was doubled [47] upon optimising the structural quality and stoichiometry. Therefore, for SECS, a much higher intrinsic $T_c$ ~ 60 − 70 K under ambient pressure can be reasonably expected in the clean bulk single crystal or thick monocrystalline film. On the other hand, $T_c$ enhancement effects under moderate hydrostatic pressure ~ 10 GPa have shown to monotonically increase (~ × 2 $T_c$) the superconducting $T_c$ of infinite-layer nickelates [40]; therefore, a high pressure enhanced superconducting $T_c \approx$ 66 K × 2 = 132 K might be achievable in this system. Considering the thickness confinement effect, oxygen off stoichiometry, and high-pressure effect, the superconducting $T_c$ of SECS can potentially mark a new milestone in superconductivity. At present, it marks a new $T_c$ record for oxides without copper under ambient pressure (**Fig. 6**).

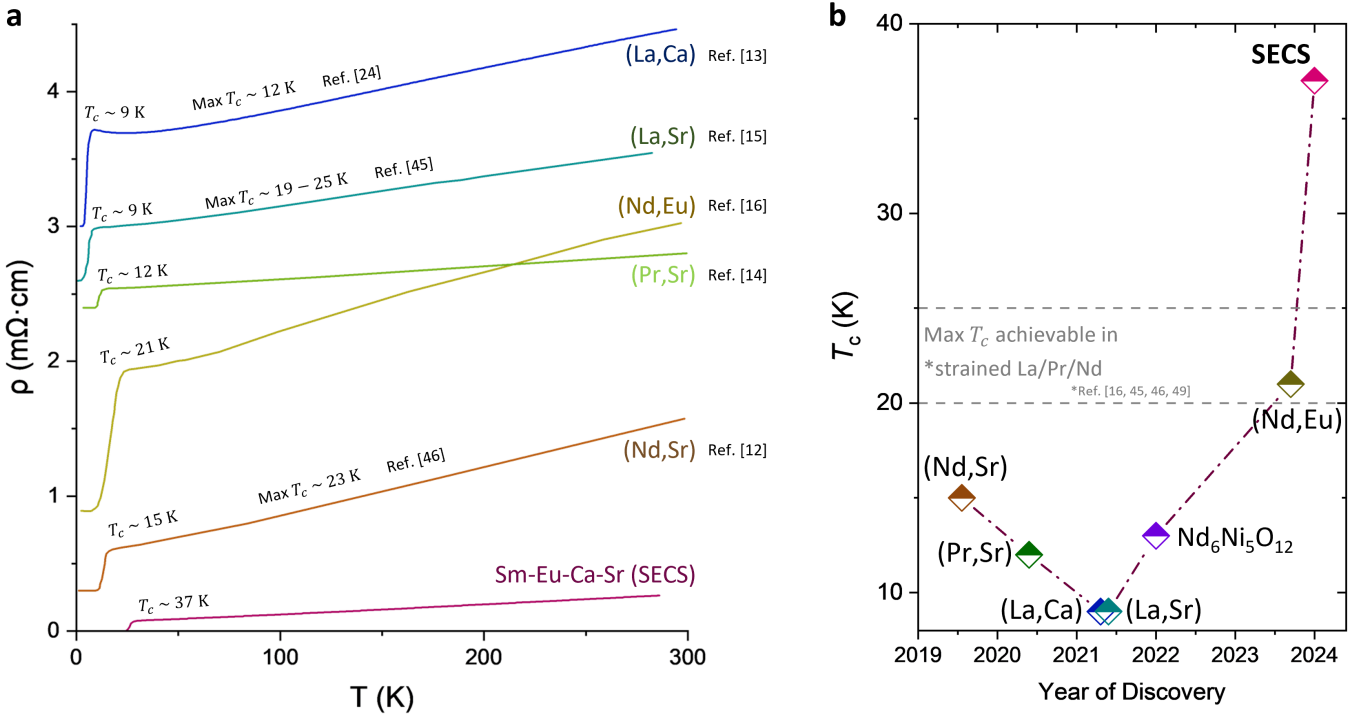

FIG. 5. Discoveries of superconducting nickel oxides. The infinite-layer nickelates family $(R, A)NiO_2$, where $R =$ La, Pr, Nd, Sm and dopant $A =$ Sr, Ca, Eu, are labelled by their A-site elements (R,A). A notable compressive strain effect might be present in $(La,A)NiO_2$//STO [13,47] and $(Nd,A)NiO_2$//LSAT [16,43,49]. (a) Resistivity-temperature $\rho(T)$ curves of the infinite-layer nickelates (first reported) [12–16]. The maximum onset of superconducting $T_c$ achieved since their first discovery is annotated. The curves are vertically shifted for clarity. (b) Superconducting $T_c$ achieved vs the year of discovery [12–17] (at ambient pressure).

In terms of nickel-based superconductivity, the present discovery of superconductivity above 30 K in SECS marks not only a new record in $T_c$ under ambient pressure but also the first successful demonstration of superconductivity in the late rare-earth nickel oxides. This step of material expansion is critical for the further understanding of the rare-earth dependence correlated orderings and pairing mechanism, which its uniqueness or universality would be the next to answer [21,56]. As a favourable ambient pressure high-$T_c$ material platform for further experimental study, this opens the doorway for particularly the low temperatures experimental studies on the superconducting phase (condition $T \ll T_c$ is more accessible to reach), such as spectroscopy study of the superconducting gap symmetry. Establishing the correlation between $T_c$ and parameters such as antiferromagnetic $J$ values may suggest a suitable interaction term for the Hubbard model description [57] and provide critical insight into the theoretical understanding of the high-$T_c$ superconducting mechanism of cuprates and nickelates.

Intriguingly, the trend in $T_c$ from the earlier rare-earth (La/Pr/Nd)NiO$_2$ to SECS is uniquely different from the expected rare earth dependence in other systems. The largest difference is typically between La$^{3+}$ and Pr$^{3+}$, Nd$^{3+}$, while the variation beyond Nd$^{3+}$ is smaller or negligible, in consistent with their ionic size variation. Using the iron pnictides as an example, the maximum bulk $T_c$ of La[O$_{1-x}$F$_x$]FeAs is ~ 26 K [58], while $T_c$ of Pr[O$_{1-x}$F$_x$]FeAs is ~ 52 K [59], and 55 K for Sm[O$_{1-x}$F$_x$]FeAs [60]. In contrast, the maximum superconducting $T_c$ of Sr-doped La-nickelate ($T_c$ ~19 − 25 K [47]) and Nd-nickelate ($T_c$ ~ 23 K [43]) are nearly identical, which makes the present observation of a much higher $T_c$ in SECS somewhat divergent from the expected rare earth dependence. However, we note that the normal state Hall coefficients $R_H$ is consistent with the expected rare earth dependence, whereas the $R_H$ sign-change temperature $T_h$ ~ 30 K for the hole-doped LaNiO$_2$ and significantly higher $T_h$ ~ 100 K for the hole-doped (Pr/Nd/Sm)NiO$_2$ including SECS. Therefore, the significantly higher superconducting $T_c$ in SECS does not correlate well with their normal state, at least in this aspect of electronic structure. On the other hand, previous experimental studies on the dimensionality of the superconducting state [42,47] of infinite-layer nickelates suggested a crossover from 2D to 3D as $T \rightarrow 0$ K that can be manipulated by the rare-earth ions and dopant size [42]. We speculate the smaller Ca$^{2+}$/Eu$^{2+/3+}$ dopants might play a role here in the dimensionality tuning of superconductivity, and consequently, a modulation in $T_c$. Such a speculation can be further verified experimentally by comparing the $T_c$ of purely Sr$^{2+}$-doped SmNiO$_2$ and purely Ca$^{2+}$/Eu$^{2+/3+}$-doped SmNiO$_2$. Nevertheless, the discovery of superconducting SECS should ignite momentum for theoretical and experimental investigations on the origin and pairing mechanism of this newfound high-$T_c$ superconductivity without copper. In principle, this report marks the first step towards direct evidence for a broad family of high-temperature unconventional superconducting oxides beyond copper.

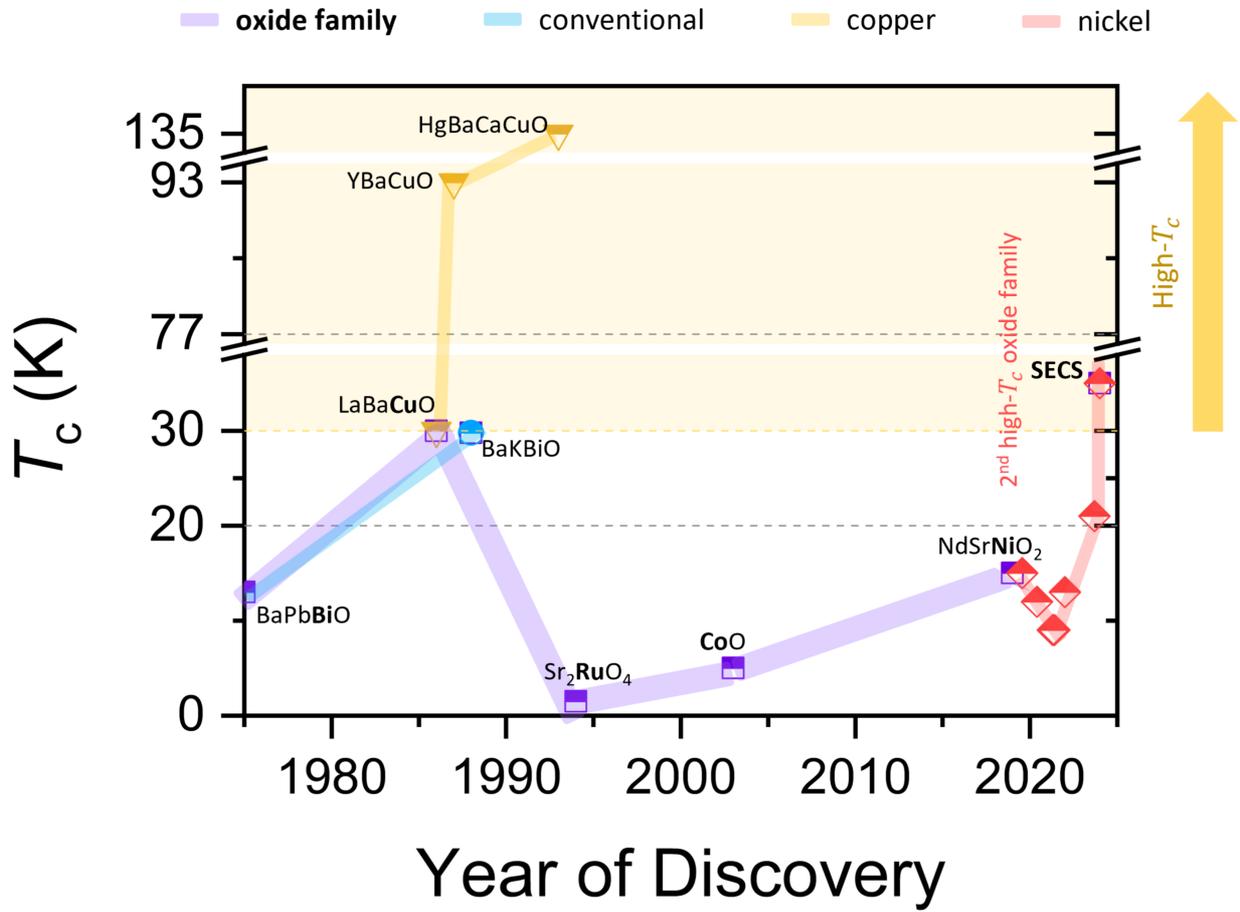

FIG. 6. Discoveries of superconducting oxides and their $T_c$ at ambient pressure.

**Acknowledgements:**

We acknowledge useful discussions with Hans J.W.M. Hilgenkamp, Dan Ferenc Segedin, Yi-Feng Yang, Yijun Yu, Bai Yang Wang, Kyuho Lee, and Wenzheng Wei. We thank the technical support from Qian He, Yang Ping, Ngee Hong Teo, Shengwei Zeng, Zhi Shiuh Lim, Saurav Prakash, and Xing Gao. This research is supported by the Ministry of Education (MOE), Singapore, under its Tier-2 Academic Research Fund (AcRF), Grant No. MOE-T2EP50121-0018 and MOE-T2EP50123-0013.

**Authors contributions:**

S.L.E.C. and A.A. conceived the project and wrote the manuscript. S.L.E.C. synthesized the nickelate films, conducted the reduction experiments, performed the transport measurements, and analysed the data. Z.L. conducted the electron microscopy.

# Supplementary Material

# for

**High-temperature Superconducting Oxide without Copper at Ambient Pressure**

S. Lin Er Chow,[†] Zhaoyang Luo, A. Ariando[†]

*Department of Physics, Faculty of Science, National University of Singapore, Singapore 117551, Singapore*

[†]Corresponding author: le.chow@u.nus.edu, ariando@nus.edu.sg

**The PDF file includes:**

Methods
Figures S1 to S9
References

## Methods

**Sample growth and preparation**

We have made over 100 trials for the synthesis of superconducting infinite-layer nickelate SECS thin films. We describe only the successful cases (and the range of parameters which lead to superconducting samples) in this manuscript. The infinite-layer nickelates Sm$_{1-x-y-z}$Eu$_x$Ca$_y$Sr$_z$NiO$_2$ (SECS) thin films were synthesized on NdGaO$_3$ (110) or NGO substrates using pulsed laser deposition (PLD) followed by topotactic reduction process in a vacuum chamber [1,2]. We note that oxygen off-stoichiometry ANiO$_{2+\delta}$ is likely in general; however, for simplicity, we annotate with the stoichiometric formula ANiO$_2$ throughout this manuscript (similarly, oxygen vacancies may form in the as grown perovskite phase ANiO$_{3-y}$; we annotate with the stoichiometric formula ANiO$_3$). NdGaO$_3$ (110) substrates (Shinkosha) were pre-annealed at 1050°C for 4 hours in air to achieve single termination surface [3]. This choice of substrate (~3.86 − 3.87 Å) is to better match the lattice constant of infinite-layer SmNiO$_2$ (~3.88 Å, the theoretical value for $P4/mmm$ space group [4]). We note that hole-doped late rare earth perovskite nickelate is particularly more challenging to synthesize than those of earlier rare earth elements or undoped perovskite, from our experience. Therefore, we started with a lower thickness of ~10 uc – 20 uc for the perovskite films, which were deposited at temperature $T_{growth} = 615°C - 625°C$ and oxygen pressure $P_{O2} = 200$ mTorr, with laser fluence of 2.15 – 2.7 Jcm$^{-2}$ at 1 Hz using a shadow mask. After deposition, samples were *in-situ* capped with epitaxial SrTiO$_3$ of ~ 4 – 7 uc (varied unintentionally due to day-to-day laser collimation fluctuations) using a low laser fluence 0.6 Jcm$^{-2}$. The reduction temperature is ~ 300°C – 325°C, with a total annealing duration of ~ 2 – 6 hours to achieve a sufficiently low resistivity below ~ 2 mΩ · cm.

**Transport characterization**

The wire connection for the electrical transport measurement was made using Al ultrasonic wire bonding. The contact resistance is typically a few hundred Ohms. The linear and Hall transport measurements at temperatures down to 2 K and magnetic fields up to 9 T were performed using a Quantum Design Physical Property Measurement System. $T_{c,0}$ is defined as the temperature at which the resistivity reaches zero (below the measurement limit ~ < 0.01 Ω or ~10$^{-7}$ Ω · cm for the few nm thin films). The onset of superconductivity is defined as the temperature where $\frac{\partial^2 \rho}{\partial T^2}$ drops below zero [5].

**Scanning Transmission Electron microscopy**

The thin film samples were first capped with gold prior the lamella preparation. A focused ion beam (FIB) instrument (FEI Versa 3D) operated at 30 kV was used to prepare cross-sectional lamellas of the nickelate thin films. Subsequently, cleaning at 2 kV was performed to remove the amorphous surface layer. The scanning transmission electron microscopy (STEM) characterization was conducted on a JEMARM200F (JEOL) microscope operated at 200 kV and equipped with a cold field emission gun and a Cs-probe aberration corrector. High angle annular dark-field (HAADF) images were acquired using inner and outer collection semi-angles of about 70 and 280 mrad, respectively, with a convergence semi-angle of about 30 mrad. Mild oxidation to the lamellas could occur during the FIB process, and over time. The imaging was done around 6 months after film reduction.

**X-ray diffraction**

The X-ray diffraction (XRD) $\theta - 2\theta$ symmetric scan was done in the X-ray Diffraction and Development (XDD) beamline at the Singapore Synchrotron Light Source (SSLS) with an X-ray wavelength of $\lambda = 1.5404$ Å.

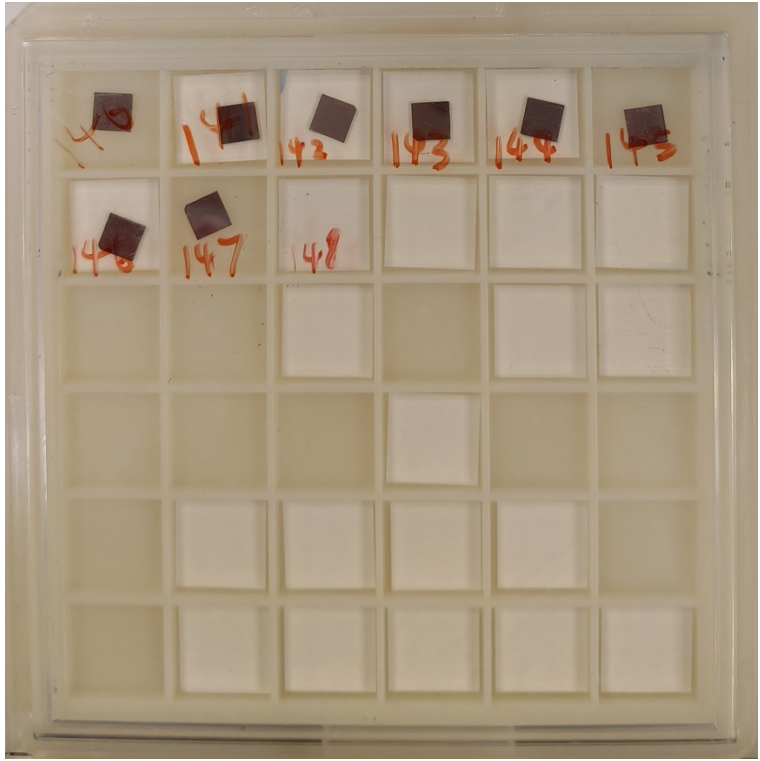

**Fig. S1.**

Photograph shows the appearance of the as grown $Sm_{1-x-y-z}Eu_xCa_ySr_zNiO_3$ thin films on the 5 × 5 mm² NGO substrates.

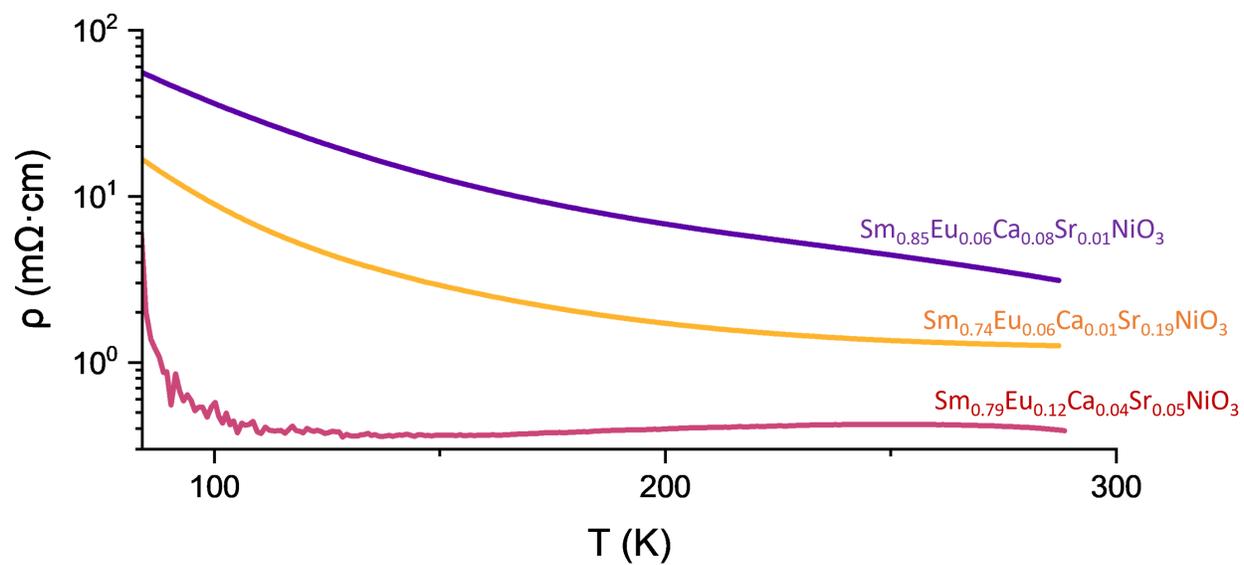

**Fig. S2.**

Resistivity *vs* temperature $\rho(T)$ curves of the perovskite $Sm_{1-x-y-z}Eu_xCa_ySr_zNiO_3$ thin films.

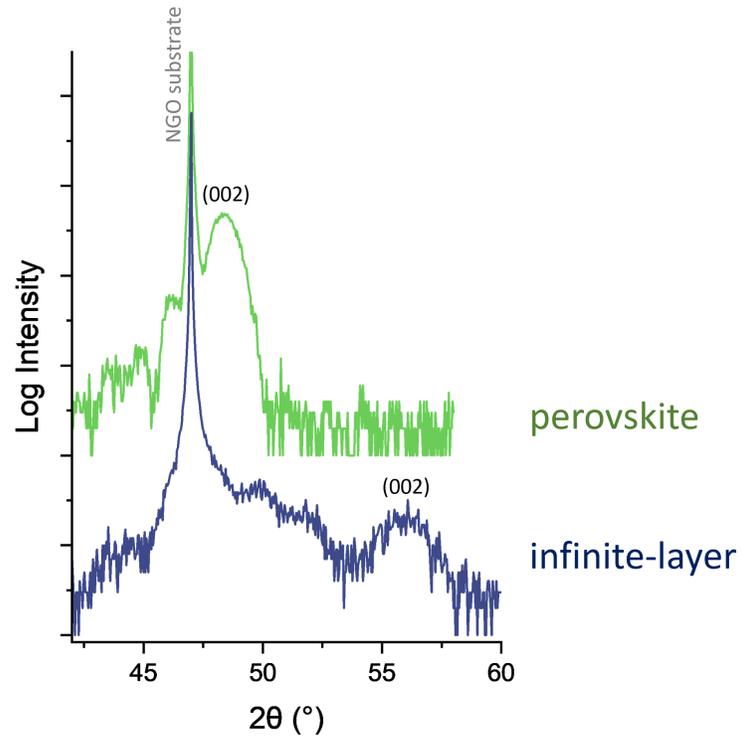

**Fig. S3.** X-ray diffraction $\theta - 2\theta$ symmetric scan of a representative sample. The calculated *c*-axis lattice constants from the (**002**) peak positions of the films are 3.76 Å for the perovskite phase and 3.29 Å for the superconducting infinite-layer phase, respectively.

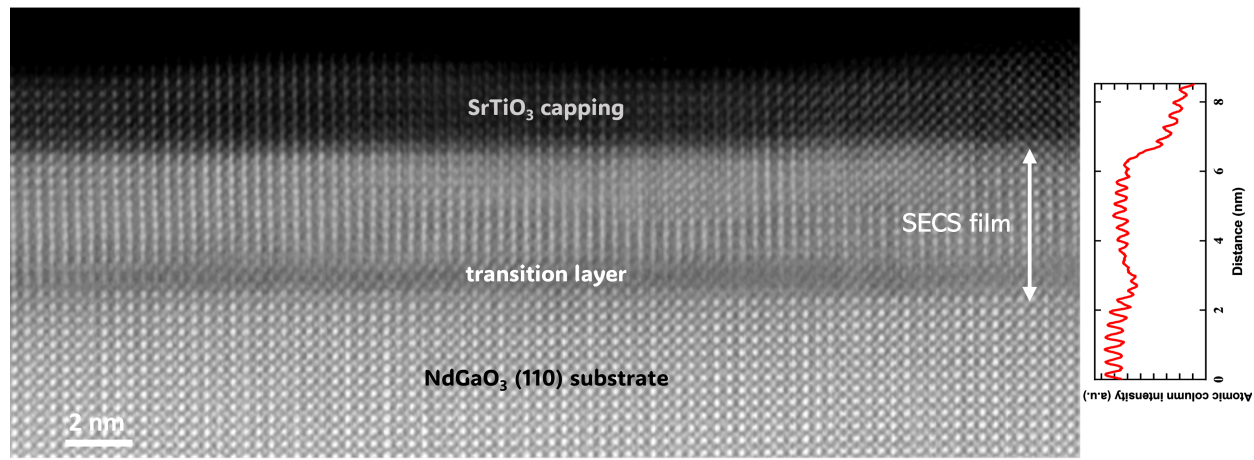

**Fig. S4.** Cross-sectional scanning transmission electron microscopy (STEM) low-magnification high-angle annular dark field (HAADF) image of a representative superconducting $Sm_{0.74}Eu_{0.06}Ca_{0.01}Sr_{0.19}NiO_2$ (SECS) thin film.

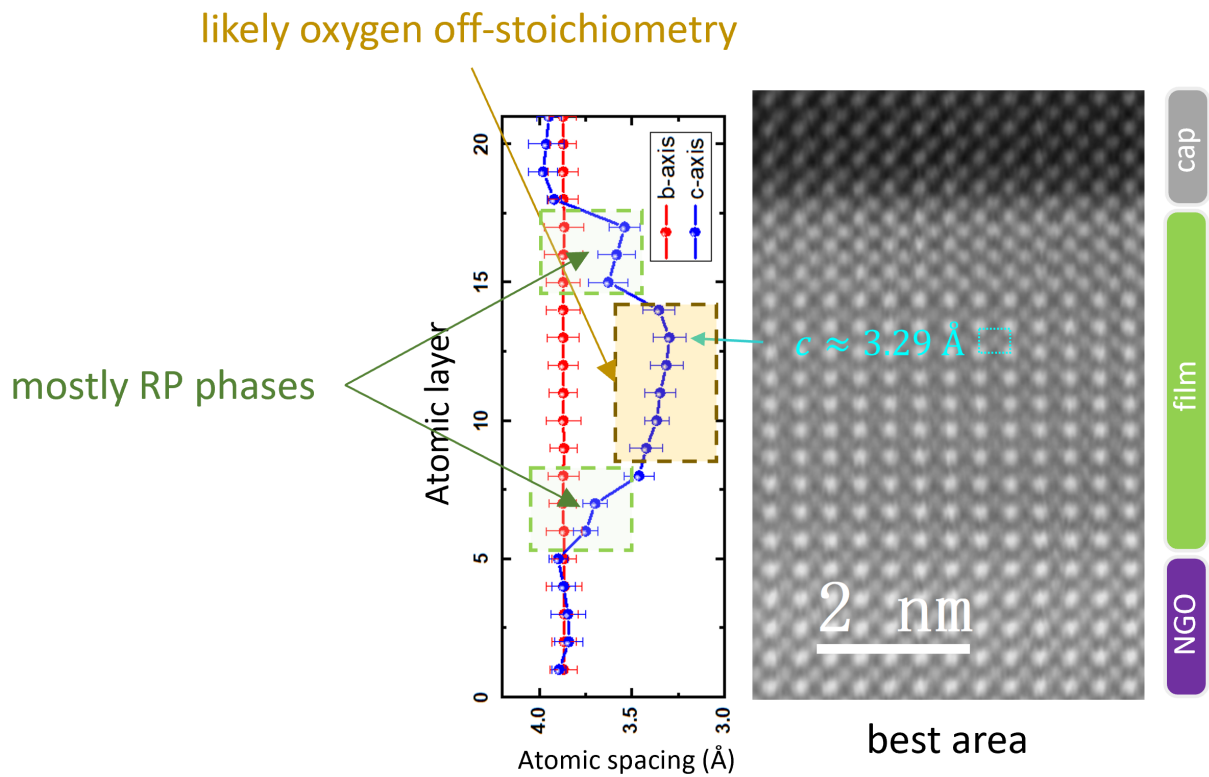

**Fig. S5.** Visualization of defects from the cross-sectional STEM-HAADF image of a representative superconducting Sm$_{0.74}$Eu$_{0.06}$Ca$_{0.01}$Sr$_{0.19}$NiO$_2$ (SECS) thin film.

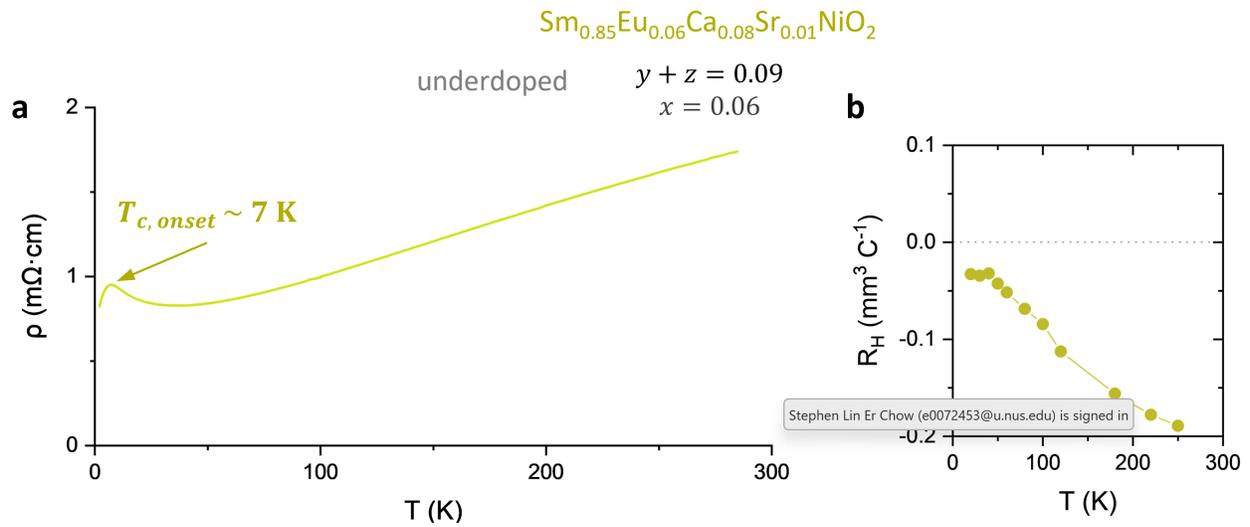

**Fig. S6.**

**Transport properties of underdoped $Sm_{0.85}Eu_{0.06}Ca_{0.08}Sr_{0.01}NiO_2$ thin films. (a)** Resistivity *vs* temperature $\rho(T)$ plot shows the onset of superconductivity at ~ 7 K and metal-insulator-transition at ~ 35 K. **(b)** Hall coefficient *vs* temperature. No crossover to positive sign was observed.

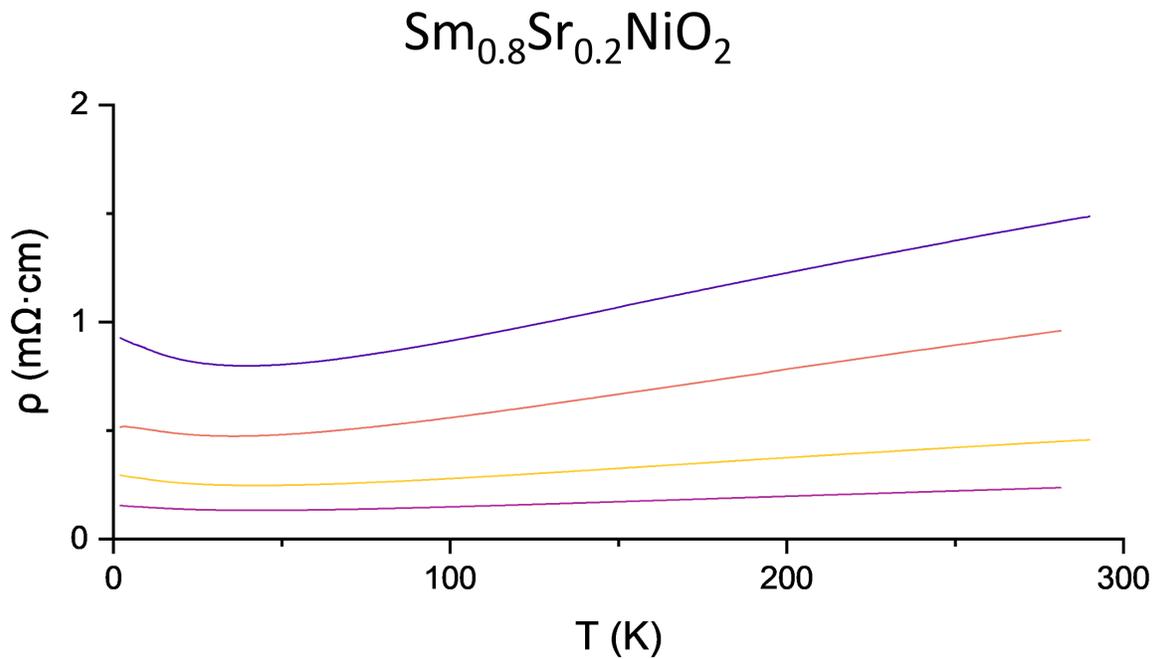

**Fig. S7.**

Resistivity *vs* temperature $\rho(T)$ curves of $Sm_{1-x-y-z}Eu_xCa_ySr_zNiO_2$ thin films with only $Sr^{2+}$ doping, $x = 0$, $y = 0$, $z = 0.2$, $Sm_{0.8}Sr_{0.2}NiO_2$. Metallic behaviour and low resistivity were observed, with a small resistivity upturn below 50 K. We emphasize that the absence of superconductivity in these samples should not be considered definitive. Considering the highly challenging synthesis, further optimisation on the growth and reduction conditions should be performed to accommodate the varied growth conditions for the different dopants.

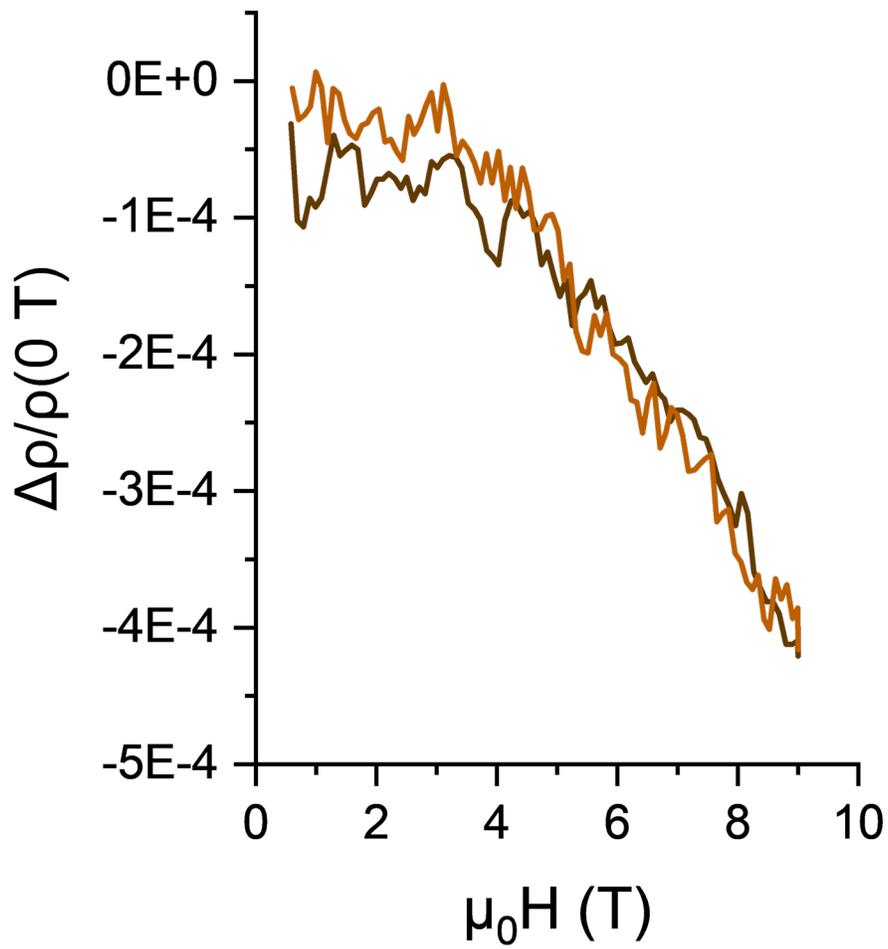

**Fig. S8.**

Magnetoresistance (MR) of two representative superconducting $Sm_{1-x-y-z}Eu_xCa_ySr_zNiO_2$ thin films at 50 K. $\Delta\rho = \rho(B) - \rho(B = 0 \text{ T})$. In all cases we observed a small negative MR, which should reflect the high defect density of the present SECS film quality.

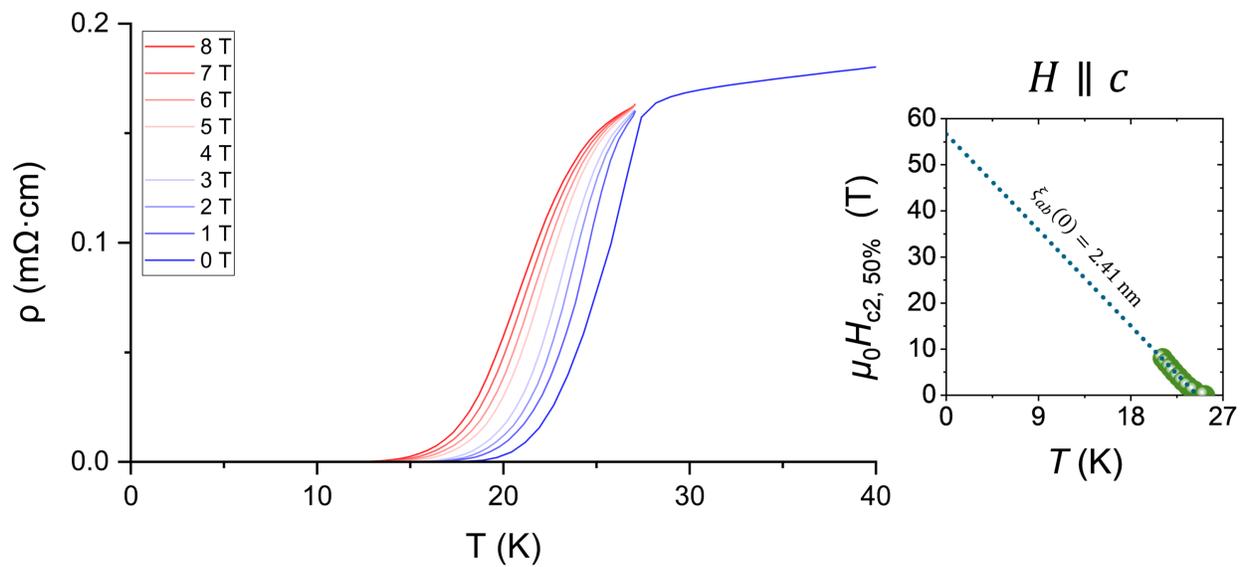

**Fig. S9.**

Magnetic-field dependence $R - T$ curve of another representative SECS sample.